\documentclass[sigplan,10pt,noacm]{acmart}

\AtBeginDocument{%
\providecommand\BibTeX{{%
\normalfont B\kern-0.5em{\scshape i\kern-0.25em b}\kern-0.8em\TeX}}}

\acmConference[]{}{}{}
\renewcommand\footnotetextcopyrightpermission[1]{}
\usepackage{amsmath}
\usepackage{verbatim} %
\usepackage{ulem}
\usepackage{subcaption} %
\usepackage{cleveref} %
\usepackage{epsfig} %
\graphicspath{{./figures/}}

\usepackage{url}

\newenvironment{tightitemize}%
{\begin{list}{$\bullet$}{%
      \setlength{\leftmargin}{10pt}
      \setlength{\itemsep}{3pt}%
      \setlength{\parsep}{0pt}%
      \setlength{\topsep}{3pt}%
      \setlength{\parskip}{0pt}%
    }%
    }%
    {\end{list}}
\newcounter{tecounter}
{\begin{list}{\arabic{tecounter}.}{%
      \usecounter{tecounter}
      \setlength{\leftmargin}{10pt}
      \setlength{\itemsep}{0pt}%
      \setlength{\parsep}{0pt}%
      \setlength{\topsep}{0pt}%
      \setlength{\parskip}{0pt}%
    }%
    }%
    {\end{list}}%

{\begin{list}{}{%
      \setlength{\leftmargin}{10pt}
      \setlength{\itemsep}{0pt}%
      \setlength{\parsep}{0pt}%
      \setlength{\topsep}{5pt}%
      \setlength{\parskip}{0pt}%
    }%
    }%
    {\end{list}}%

\crefformat{section}{\S#2#1#3} %
\crefformat{subsection}{\S#2#1#3}
\crefformat{subsubsection}{\S#2#1#3}

\usepackage{tabularx}
\newcounter{protocol}
\newenvironment{protocol}[1]
{\par\addvspace{\topsep}
  \noindent
  \tabularx{\linewidth}{@{}
    X @{}} \hline \refstepcounter{protocol}\textbf{Protocol \theprotocol} #1 \\
  \hline} { \\ \hline \endtabularx \par\addvspace{\topsep}}
\newcommand{\sbline}{\\[.5\normalbaselineskip]} 

\begin{document} 

\date{} 

\newcommand{\sys}{\textbf{TEE/GC Hybrid}} \newcommand{\shortsys}{\textbf{TGh}}
\title{\shortsys: A TEE/GC Hybrid Enabling Confidential FaaS Platforms }

\author{ James Choncholas }
\affiliation{Georgia Institute of Technology}
\email{jgc@gatech.edu}

\author{ Ketan Bhardwaj }
\affiliation{Georgia Institute of Technology}
\email{ketanbj@gatech.edu}

\author{ Ada Gavrilovska }
\affiliation{Georgia Institute of Technology}
\email{ada@cc.gatech.edu}

\maketitle 

\subsection*{Abstract} Trusted Execution Environments (TEEs) suffer from
performance issues when executing certain management instructions, such as
creating an enclave, context switching in and out of protected mode, and
swapping cached pages.
This is especially problematic for short-running, interactive functions in
Function-as-a-Service (FaaS) platforms, where existing techniques to address
enclave overheads are insufficient.
We find FaaS functions can spend more time managing the enclave than executing
application instructions.
In this work, we propose a TEE/GC hybrid (TGh) protocol to enable confidential
FaaS platforms.
TGh moves computation out of the enclave onto the untrusted host using garbled
circuits (GC), a cryptographic construction for secure function evaluation.
Our approach retains the security guarantees of enclaves while avoiding the
performance issues associated with enclave management instructions.

\section{Introduction}
\label{sec:intro}
Software and data protected within a Trusted Execution Environment are isolated
from a compromised OS, malicious userspace processes, and other malicious TEEs
when operating as intended, a promise of confidential computing.
To fully capture the benefits of TEEs, recent work in
industry~\cite{openenclave, asylo} and academia~\cite{graphene,scone,occlum}
has incorporated these hardware-based security features into systems which make
them efficient, easy to consume, and easy to manage.

A common challenge for these systems is to amortize the overheads of TEE-based
execution.
These overheads stem from managing hardware structures when creating the
enclave, switching context, and accessing memory which does not fit within the
enclave page cache~\cite{hotcalls}.
Such overheads are reasonable over the lifespan of long-running tasks using
tricks to batch and reorder I/O, Intel's Switchless calls, and reducing enclave
memory usage~\cite{hotcalls,sgxperf}.

However, in the context of short running and interactive tasks such as in
Function-as-a-Service (FaaS), the overhead of trusted execution is quite
high~\cite{faas-confidential}.
We observe that the BeFaaS benchmark~\cite{befaas} contains only a small
handful of operations per function, a much smaller cost than the 17,000 cycles
required just to perform the ecall to pass data into the
enclave~\cite{hotcalls}.
Existing approaches to address TEE overhead, like HotCalls and Intel's
Switchless Calls, do not fix the fundamental issue for short interactive tasks
which, by definition, require frequent context switches for I/O and fast
starts.

\begin{figure}[t]
  \centering
  \includegraphics[width=\columnwidth]{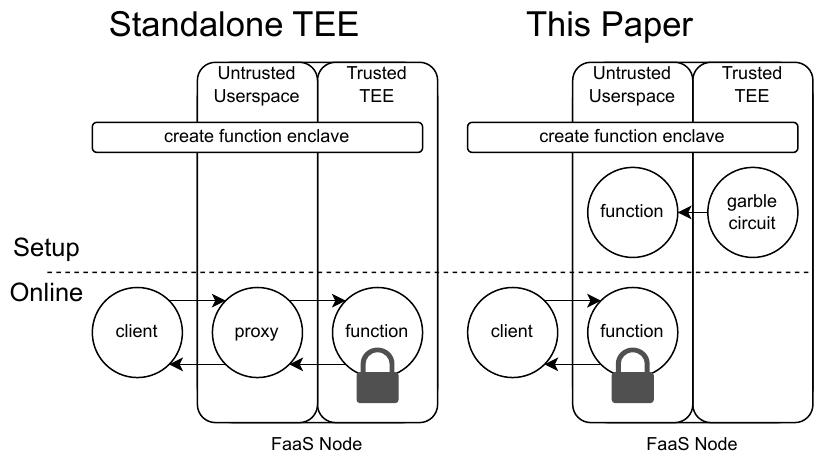}
  \vspace{-5ex}
  \caption{Comparison of confidential computing techniques.}
  \label{fig:overview}
  \vspace{-2ex}
\end{figure}

In this work, we propose a rather unorthodox approach for trusted execution
which we call \sys{}.
The idea is to bypass TEE hardware inefficiencies by repurposing techniques
from Secure Multiparty Computation (MPC).
MPC is a field of cryptography used to securely evaluate functions between two
mutually distrusting parties (often on physical separate machines), each
wanting to compute on the aggregate of their secret data without sharing the
data with each other.
We re-purpose a specific MPC construction, garbled circuits (GC) to run between
enclave and host (on same physical machine), while retaining the same security
guarantees as native execution on the TEE.
The key idea is the trusted enclave sets up a function for the untrusted host
to evaluate.
The host then evaluates the function in userspace without the performance
penalties of context switching or the overhead of enclave page cache
management, as depicted in~\autoref{fig:overview}.

MPC protocols have seen dramatic theoretical improvements in efficiency over
the last decade, however security is not without cost.
Secure function evaluation under MPC is orders of magnitude slower than
evaluating the same function natively.
Intuitively, this would preclude MPC protocols from being useful compared to
hardware enclaves, however, we notice that when MPC is used to offload
computation in this setting, many simplifications can be made to the protocols.
In general, MPC enables collaborative computation between many parties each of
which may have secret data.
Computation offload on the other hand is a subset of this scenario where only
one party (the enclave in this case) has secret data.
We apply garbled circuits to this problem such that expensive cryptographic
operations like Oblivious Transfer (OT)~\cite{ot} are unnecessary as only one
party owns all the secret data.
As such, GC is simpler and less expensive in the configuration we propose for
two reasons: the lack of OT, and colocation of the enclave and host who
evaluate the protocol among each other.
Using the EMP toolkit library, we measure GC evaluation speed over a LAN to be
5 million AND gates per second, and increases from 22 million to 35 million
simply from transferring the garbled truth tables over shared memory vs.
\ local
loopback.
For short running functions this makes it possible to evaluate their GC faster
than the ecall into the TEE, making a TEE/GC hybrid (\shortsys{}) approach an
enabler for confidential FaaS.
An important limitation of \shortsys{} is the constant cryptographic overhead
of evaluating garbled circuits.
Since every operation under GC is slower except for enclave management related
operations like ecalls and EPC evictions, only short running functions benefit
from this approach.
Furthermore, functions running under \shortsys{} need to be reimplemented as
boolean circuits such that control flow does not depend on secret data.
Not all functions are ammenable to this transformation.

The remainder of this paper describes our research contribution.
We detail a TEE/MPC hybrid approach to trusted execution, supplemented with
experiments motivating the envisioned performance properties.
\vspace{-2ex}
\section{Background}
\label{sec:prelim}
\sys{} consists of two fundamental technologies, Trusted Execution Environments
and garbled circuits.

\noindent {\bf TEE. }
The two major processor designers each have their own implementation of a TEE,
Intel with Software Guard Extensions (SGX) and ARM with TrustZone.
The set of hardware features collectively called the TEE provide an elevated
degree of security for applications.
SGX specifically extends the x86-64 ISA to allow application to instantiate a
protected execution environment called an enclave while only trusting the
hardware and not system software (hypervisor, OS, frameworks, etc.) with
explicit instructions to perform host to enclave switch (ecall) and vice versa
(ocall).
It also incorporates memory protection.
When executing in enclave mode the processor enforces additional checks on
memory accesses ensuring that only code inside the enclave can access its own
enclave region.
For other features and details readers are directed to the SGX and TrustZone
specifications~\cite{sgxmanual,trustzonemanual}.

\noindent {\bf GC. }
Garbled circuits were invented by Andrew Yao in 1986~\cite{yao}.
They are a cryptographic construction to enable secure function evaluation,
allowing multiple parties to compute on the aggregate of their secret data
without revealing it to one another.
Most basically, garbled circuits is a two party protocol with one party playing
the role of the generator and one of the evaluator.
A garbled truth table is the foundational component of a garbled circuit,
generated by the generator and passed to the evaluator.
A garbled truth table is a set of random strings designated as outputs
encrypted by sets of random strings designated as inputs.
The encrypted outputs can then be used as subsequent inputs to other garbled
truth tables allowing generic secure function evaluation.
This allows multiple parties to securely compute a function over their inputs
without revealing anything but the output.

\vspace{-2ex}
\section{Feasibility and Challenges}
\label{sec:design}
A high level overview of \sys{} enabled confidential computing is depicted in
\autoref{fig:overview}.
The key to our approach is to offload computation from the enclave to an
untrusted process running on the host in such a way that retains enclave
security guarantees.
The process running outside the enclave is evaluating a garbled circuit and is
not subject to ecall/ocall overheads, memory access overheads, and cannot see
plaintext data on which it is computing.
Given the high overheads of cryptography-based secure function evaluation, it
is important to ensure a performance is a closely monitored for this new
setting where MPC is used to offload computation.

\begin{figure}
  \begin{protocol}{TEE/GC Hybrid Scheme}
    \sbline
    \textit{Inputs.}
    TEE~$T$ holds input~$x$ and would like to offload computation of the function
    $f(x)$, represented as a boolean circuit, to the untrusted host $H$.

    \sbline
    \textit{Goal.}
    Host~$H$ computes $y=f(x)$ without learning anything about $x$ or $y$.

    \sbline
    \textit{Protocol:}
    \begin{enumerate}
      \item \textbf{Setup phase.}
            Before the input to the function is known, $T$ may begin by generating a
            standard garbled circuit.
            We present this in the notation of~\cite{wrk}.
            \begin{enumerate}
              \item
                    $T$ associates a random mask bit $\lambda_\alpha \in
                      \{0,1\}$ with
                    every wire of the circuit, enabling the point-and-permute
                    technique of~\cite{pointandpermute}.
                    $T$ also associates random labels $\lambda_{\alpha, 0}$ and
                    $\lambda_{\alpha, 1} = \lambda_{\alpha, 0} \oplus \Delta$
                    with every wire, enabling the free-XOR technique
                    of~\cite{freexor}.

              \item
                    $T$ generates a garbled truth table of the form:
                    \begin{tabular}{llll}
                      \hline
                      $\hat{x}$ & $\hat{y}$ & garbled rows
                      \\
                      \hline
                      0         & 0         & $H(L_{\alpha,0}, L_{\beta,0},
                        \gamma, 00)
                        \oplus (\hat{z}_{0,0}, L_{\gamma, \hat{z}_{0,0}})$
                      \\
                      0         & 1         & $H(L_{\alpha,0}, L_{\beta,1},
                        \gamma, 01)
                        \oplus (\hat{z}_{0,0}, L_{\gamma, \hat{z}_{0,1}})$
                      \\
                      1         & 0         & $H(L_{\alpha,1}, L_{\beta,0},
                        \gamma, 10)
                        \oplus (\hat{z}_{0,0}, L_{\gamma, \hat{z}_{1,0}})$
                      \\
                      1         & 1         & $H(L_{\alpha,1}, L_{\beta,1},
                        \gamma, 11)
                        \oplus (\hat{z}_{0,0}, L_{\gamma, \hat{z}_{1,1}})$
                      \\
                      \hline
                    \end{tabular}
                    where $H()$ is a hash function modeled as a random oracle.
                    AES is commonly used in practice.
                    The labels may be chosen pseudorandomly using the output of a PRF.

              \item
                    $T$ sends all garbled truth tables to $H$ through shared
                    unencrypted
                    memory pages.
                    Thus, $H$ learns the masked bits $\hat{x} = x \oplus \lambda_\alpha$ and
                    $\hat{y}$, as well as the garbled truth table.

            \end{enumerate}
      \item \textbf{Online phase.}
            \begin{enumerate}
              \item
                    For every input wire of the circuit, $T$ sends $H$ one of
                    the two
                    possible
                    wire labels per gate.
                    Alternatively, a remote client who has deployed enclave $T$ may send the wire
                    labels to $H$ directly.
                    This is easy as the remote client would know seed $S$ and be able to generate
                    the wire labels without requiring any interaction with $T$.

              \item
                    $H$ evaluates the garbled circuit as usual.
                    For AND gates, the row indexed by $\hat{x}$, $\hat{y}$ is decrypted yielding
                    $\hat{z}$ and $L_{\gamma, \hat{z}_{0,0}}$.
                    XOR gate labels may simply be XOR'ed together as described by~\cite{freexor}

              \item Upon reaching output gates in the circuit, $H$ sends the output labels to
                    whomever must learn the output, either $T$ or a remote client.
                    If the receiver recognizes the labels received from $H$ as labels assigned to
                    output wires, it learns the plaintext result of the computation.
                    If the receiver does not recognize the labels received, the circuit was not
                    correctly evaluated and is aborted.
            \end{enumerate}
    \end{enumerate}

  \end{protocol}
  \label{protocol:teegc}
\end{figure}

\noindent {\bf TEE/GC Hybrid Protocol: }
We first give a high level intuition of the TEE/GC hybrid protocol, then
discuss how it differs from standard garbled circuits.
To offload computation, the enclave first generates a garbled circuit, a
process which in practice mostly consists of repeatedly evaluating a block
cipher e.g. AES.
The garbled circuit can be precomputed and shared before the input is known.
The circuit is then sent to an untrusted process on the host through pages
mapped into the address spaces of both enclave and host process.
Note this memory is not encrypted, it is only used to transfer the garbled
circuit.
When the enclave or a remote client wishes to evaluate a function, it sends
wire labels associated with their input to the host.
The untrusted host evaluates the garbled circuit gate by gate, a process which
again mostly involves evaluating AES in practice.
The host then responds with the labels associated with output wires.
This process is formally stated in Protocol 1.

While we haven't changed the core of how garbled circuits work, there are a few
key differences from how garbled circuits typically are used.
Most obviously, all secret inputs are owned by the TEE in the scenario we
consider.
This means that the TEE can directly send wire labels to the untrusted host.
In garbled circuits, typically both parties have inputs and wire labels
corresponding to the circuit evaluator's inputs need to be transferred using an
expensive cryptographic primitive, Oblivious Transfer.
While this is hardly a surprise to those familiar with garbled circuits, it is
important to notice how the parties are configured.
Oblivious Transfer is an expensive primitive used in MPC to send secret inputs
to a function.
The more interesting difference between Protocol 1 and standard garbled
circuits is Protocol 1 explicitly states the enclave uses a seeded pseudorandom
function PRF to generate wire masks and labels for the garbled circuit.
While this is common in practice~(\cite{emp-toolkit}), in this context it
allows the enclave and remote client to share the PRF seed in the setup phase.
Then, in the online phase, the client may generate the same garbled labels that
the enclave used to create the circuit and directly send the wire labels to the
untrusted host without interacting with the enclave, thereby moving the enclave
out of the critical path.

In \shortsys{}, the only purpose of the enclave is to generate the correlated
randomness later used by the host to evaluate the garbled circuit.
As such, only a single enclave per machine is required as it may be shared by
all clients.
As long as clients are convinced of the integrity of the code running inside
the enclave via attestation, a single enclave may generate the correlated
randomness using a unique seed per client.
\vspace{-2ex}
\section{Preliminary Evaluation}

\noindent {\bf Security.}
The security of our TEE/GC hybrid falls to the lowest common denominator of
TEEs and GC.
Specifically, the hybrid scheme is broken if either the garbled circuit is
broken or the TEE is compromised.
This is good, as it means the hybrid scheme is just as secure as a TEE.

We prove this by contractiction, namely if an attacker can break the TEE/GC
hybrid, they have broken either the garbled circuit or the TEE.
Say the untrusted host attacking Protocol 1 learns the plaintext value of an
intermediate wire label in the computation beyond a negligible advantage
(better than flipping a coin.)
Considering the view of the host contains only garbled truth tables, this
implies that the host has either broken the garbled circuit and can reverse the
block cipher used to generate the garbled circuit with non-negligible
probability, or the host has learned this information out-of-band from the TEE.
TEE security assumptions, however, are a superset of garbled circuits.
Both assume block ciphers act as random oracles; TEEs encrypt RAM with AES and
garbled circuits build truth tables with it.
However, TEEs have a litany of other cryptographic
assumptions~\cite{sgxmanual}.
Since TEEs are strictly weaker than garbled circuits, TEE/GC hybrid has similar
security properties as TEEs alone and security guarantees have not been
weakened by introducing garbled circuits.

\noindent {\bf Performance. }
Evaluating a garbled circuit outside an enclave is slower than plaintext
execution inside an enclave, however the GC does not need to start an enclave,
ecall, ocall, or page.
Thus, the crux of the performance question we explore is the following: out of
a set of instructions, how many must be related to enclave management
operations to warrant offloading via \shortsys?
The more management operations which exists in a set of instructions, the
higher the overhead of the TEE, overhead that can be eliminated by executing
those instructions as a GC outside the enclave.

Enclave management operations consist of the following: \begin{tightitemize}
  \item ecalls and ocalls which may transfer buffers.
  \item Writing to encrypted and unencrypted memory.
  \item Initializing and destroying the enclave.
\end{tightitemize}

This work focuses on operations in the critical path of FaaS-based systems and
operations which are known a priori to be cause for concern.
Specifically we consider ecalls (ocall performance is similar enough for this
analysis), EPC page evictions, and enclave initialization.
In the remainder of this section we compare enclave operations to the number of
AND gates which can be evaluated per second.
These numbers were measured from the EMP~\cite{emp-toolkit} library running on
a Intel Core i9 11th generation CPU, measuring the rate at which EMP can
evaluate garbled truth tables.
Our results focus on AND gates as XOR gates can be evaluated without AES and
thus are much faster.

\begin{figure}[t]
  \centering
  \includegraphics[width=\columnwidth]{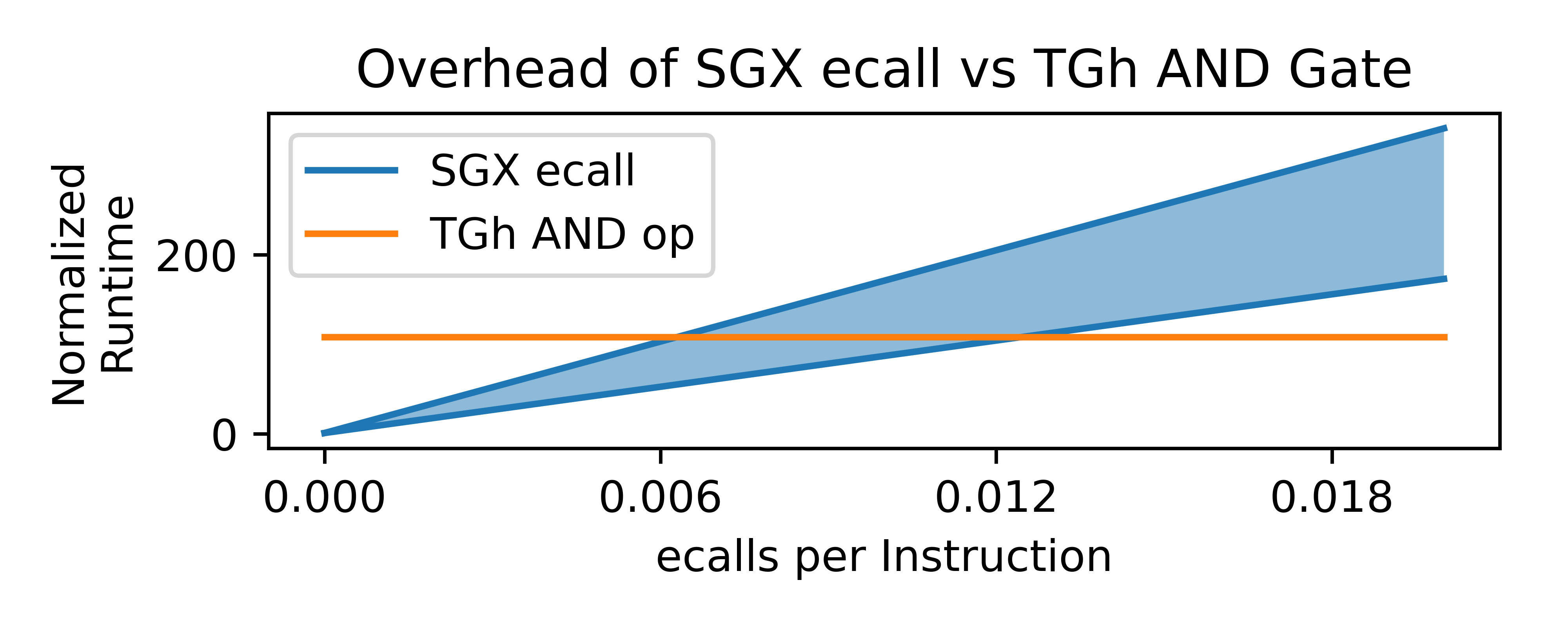}
  \vspace{-4ex}
  \caption{\sys{} can evaluate one AND gate in the same time as a set of
    instructions made up of $1\%$ ecalls (assuming non-ecall instructions
    execute in $1$ cycle).}
  \label{fig:ecalls}
  \vspace{-2ex}
\end{figure}

\noindent {\it Bypassing ecalls overhead:}
According to recent work~\cite{hotcalls}, ecalls on Intel SGX can consume up to
17,000 extra cycles.
This includes the direct effects of context switching like flushing caches and
TLBs, but it also includes indirect costs of subsequent compulsory cache and
TLB misses.
They also measure a minimum costs of 8,600 cycles while other work claims
ecalls cost a similar 10,000 cycles~\cite{sgxvirt}.
The upper and lower bounds are shown in \autoref{fig:ecalls} compared to the
constant amount of time it takes to evaluate one AND gate using \shortsys{},
outside the enclave.

The important inflection point in \autoref{fig:ecalls} is for every $0.7\%$ of
ecalls that a series of instructions contains, one garbled AND gate can be
evaluated in the same amount of time, outside the enclave.
Thus, simple functions which can be represented in a small number of AND gates
can theoretically run faster as a garbled circuit with increasing benefit as
the program requires more ecalls.
In practice, however, functions rarely consist of such a small number of AND
gates, thus, ecall overhead is alone is not enough to justify the high overhead
of garbled circuits.

  {\it Bypassing EPC eviction overhead:}
Intel SGX has an enclave page cache (EPC) to store metadata about encrypted
pages.
When the number of pages grows beyond what this data structure can track, pages
must be evicted through an expensive process that involves multiple memory
accesses to encrypted data.
In Intel's SGXv1 the EPC is either 128MB or 256MB while in SGXv2, it can be up
to 512GiB per socket.
SGXv2 is a huge step towards enabling enclaves for applications with a large
working set of memory, however in multitenant situations such as in FaaS, even
the large EPC size may pale in comparison to the maximum amount of DRAM such a
machine could be configured with (and legitimately need).
The EPC is shared across all enclaves raising questions of performance
isolation between tenants.
Thus, even with the massive increase in EPC memory size in SGXv2, we still
consider the performance implications of EPC page evictions due to the concerns
with multitenancy and the fact that SGXv2 is now only available on Xeon
server-grade SKUs and support has been dropped for Core series consumer grade
processors.
According to Ngoc et al.~\cite{sgxvirt}, one EPC page eviction consumes 12,000
cycles.

\begin{figure}[t]
  \centering
  \includegraphics[width=\columnwidth]{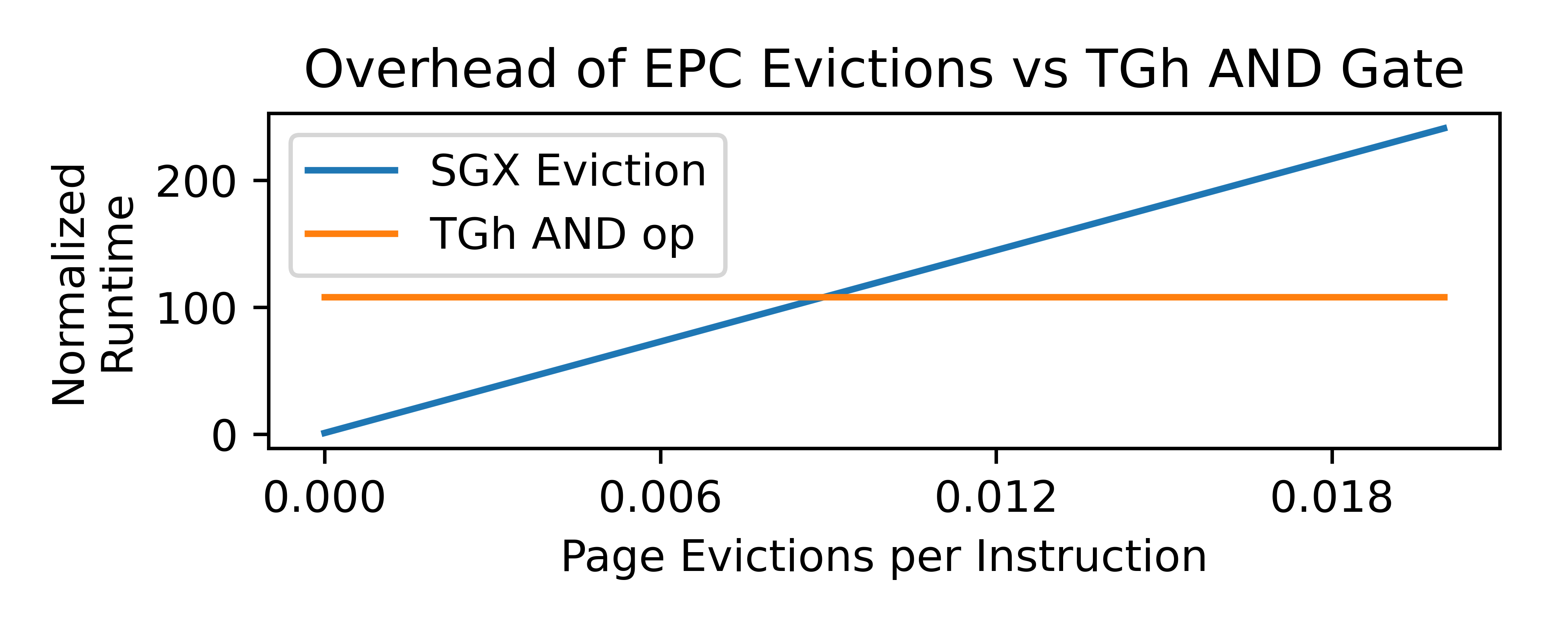}
  \vspace{-5ex}
  \caption{\sys{} can evaluate one AND gate in the same time as a set of
    instructions with $0.8\%$ causing EPC page evictions (assuming non-evicting
    instructions execute in $1$ cycle). }
  \vspace{-2ex}
  \label{fig:epc}
\end{figure}

In \autoref{fig:epc} we can see the inflection point of when EPC page eviction
cost outweighs the cost of garble circuits.
For every $0.8\%$ of instructions which cause an EPC page eviction, one garbled
AND gate can be evaluated in the same amount of time.
Thus, a set of instructions causing frequent EPC page evictions runs faster
using \shortsys{} if the function can be represented in a small number of AND
gates.
However, given recent increase in EPC size from 128MB in SGXv1 to 256GB in
SGXv2, it is unlikely that the cost of evictions alone will outweigh the
overhead of executing a function as a garbled circuit.
As such, EPC page evictions are complimentary to more dramatic performance
reasons for \shortsys{}, such as enclave creation.

  {\it Bypassing enclave creation overhead:}
According to Gjerdrum et al.~\cite{sgxstartupslow}, it takes 300ms to create a
batch of 100 SGX enclaves.
The relationship between creation time and batch size is linear thus we can
expect creation of a single enclave to take 3 milliseconds.
In the same amount of time 111000 AND gates can be evaluated, or AES can be
evaluated under MPC 17 times.
Furthermore, LibOS-based approaches to TEE programming further exaserbate
startup costs with simple no-op (return 0;) TEE calls requiring 300 ecalls,
1000 ocalls, and 1000 AEX exits measured on SGXv1~\cite{sgxgauge}, and taking
$370ms$ measured on SGXv2~\cite{sgxgauge,clemmys}.
Thus, removing enclave creation from the critical path leaves room to run
simple functions at the higher cost of garbled circuit evaluation.

\noindent {\bf End to End Performance in FaaS benchmark: }
Thus far we have presented the cost of enclave overheads individually compared
to garbled circuits, but how do these overheads stack up in a real application?
To give real-world context we consider the BeFaaS e-commerce application which
is based on Google’s microservice demo~\cite{befaas}.
BeFaaS is a collection of functions which implement an online shopping app, one
such function allowing a user to check out.
To synthetically compare this to \shortsys{}, we represent the checkout
function as a Boolean circuit, then count the number of AND gates to project
how long it would take to execute as a GC.
With 2488 AND gates, computing the checkout function is projected to take
$70$us under GC.
This number is significant because it is lower than enclave startup time,
faster than doing ~20 ecalls, and faster than ~25 EPC page evictions.

Looking beyond applications, we note the comparisons we've drawn to garbled
circuits come from numbers gathered using the EMP library.
Semi-honest garbled circuit evaluation in EMP (the specific implementation
we're using) is single threaded.
In the GC literature, the dominant cost is network bandwidth not CPU
computation, thus there is no benefit to parallelizing circuit evaluation.
In the context of \sys{}, however, the network between the parties is instead a
much higher bandwidth using shared memory pages.
As such, parallelizing circuit evaluation can significant improve on GC
evaluation speed, which is why we refer to \sys{} as being accelerator
friendly.
\vspace{-2ex}
\section{Discussion}

\noindent {\bf Benefits: }
One benefit of \shortsys{} is the simplicity of the software which runs in the
enclave.
Since all the enclave does is generate garbled circuits, the interface between
the untrusted host and enclave is thin minimizing the attack surface of the
enclave code.
In \shortsys{}, the enclave is basically acting as an oblivious PRF which can
be evaluated inexpensively but is subject to side-channel leakages compared to
pure cryptographic approaches.

Another benefit is that multiple mutually distrusting remote clients may use
the same enclave, removing enclave creation time from the critical path.
Since all the enclave does is generate garbled circuits, the enclave source
code may be available for all to inspect, and clients who trust the attestation
of enclave authenticity can be confident that sharing an enclave to generate
garbled circuits will not reveal their private data.
As such, the enclave may act as an orchestrator for many remote clients,
evaluating their tasks on the host machine.
Each client does not need to pay the cost of enclave creation, nor deal with
issues of performance isolation as all enclaves running on the same host must
share the protected enclave page cache (EPC).
Low complexity software running inside the enclave makes it easier for multiple
clients to audit and trust there are no bugs in the enclave circuit generation
code.

\sys{} also achieves active security without supplemental cryptographic
techniques like message authentication codes commonly used to improve security
guarantees of other protocols~\cite{bdoz,spdz,wrk}.
What this means is the enclave (or the remote client) can tell by looking at
the output from the host if the host tried to cheat in the protocol.
The only messages the host sends (and thus can cheat on) are the output wire
labels.
The host will obtain at most one out of two output labels, by the nature of
garbled circuits, and the only way for the host to obtain the one label is to
correctly evaluate all gates up until the output.
Thus, if the host does not correctly evaluate the garbled circuit they will not
receive a legitimate label on the output wires.
If the receiver (enclave or remote client) receives an invalid label from the
host, they know the host has not correctly evaluated the circuit.
This shows the only opportunity the host has to cheat is to guess the output
wire label it did not learn with probability of success $1/2^\sigma$, with the
random oracle assumption and $\sigma$ being a statistical security parameter.

Lastly, we would like to note that garbled circuits are friendly to hardware
acceleration.
Recent work have accelerated garbled circuits using everything from
GPUs~\cite{gpugc} to ASICs~\cite{asicgc}.
Garbled circuits parallelize to the same degree as the underlying function and
mostly consist of repeatedly evaluating AES.

\noindent{\bf Limitations: }
TEEs are subject to side channel attacks unlike MPC
protocols~\cite{sgxprobeattack,sgaxe,sgxforeshadow}.
One might assume combining the TEEs and MPC may improve security beyond what
each offers alone, however this is not the case.
Instead, security guarantees fall to the lowest common denominator, however as
we show, our TEE/MPC hybrid is no weaker than TEEs alone.
As proposed, MPC is leveraged to improve the performance of TEE execution, not
TEE security guarantees but future TEE/MPC hybrids may extend beyond
performance to security.

Furthermore, \shortsys{} does not hide data access patterns, a goal of recent
work using TEEs to build Oblivious RAM (ORAM) schemes~\cite{obaldi}.
Additionally, certain MPC protocols like those based on garbled circuits are
expensive for tasks with branch-y control flow.
While recent efforts address the cost of branches under MPC~\cite{stackedgc},
it has historically required data oblivious algorithms and predicated
execution.

Secondly, executing a function as a garbled circuit cannot easily be done with
an unmodified binary, as with LibOS based approaches discussed later.
Thus \shortsys{} requires additional engineering effort to build functions as
circuits to be executed outside the enclave.
Furthermore, GC require data oblivious algorithms and predicated execution to
prevent leaking private data through conditionals.
This leads to more engineering, and potentially performance degradation in
building functions as circuits.

\noindent {\bf Related work.}
A popular approach to using enclaves is via library OS which supports running
an unmodified application binary within an enclave and using the dynamic linker
to capture system calls, which are redirected to the host OS.
Prior work reports running an empty enclave (return 0;) on one such system,
Graphene, to perform 300 ecalls, 1000 ocalls, 1000 AEX exits, and 1M EPC
evictions~\cite{sgxgauge}.
This was measured on SGXv1 which does not support dynamic memory allocation,
thus the entire default sized 4GiB heap must be preallocated and paged out
which explains the high number of EPC evictions.
While SGXv2 does support dymanic allocation and thus does not have this high
EPC cost, SGXv2-based platforms still see slow enclave creation time e.g.
370ms~\cite{clemmys}.
In the same amount of time, over 13 million AND gates can be evaluated
corresponding to evaluating AES under MPC 2140 times.
Being able to run unmodified binaries within TEEs greatly reduces development
overhead but comes at the cost of performance, an especially high cost for
short running tasks.
Frequently paying such cost, for example on function cold starts, highlights
the usefullness of our approach.

\noindent {\bf Open Questions.}
\shortsys{} is a novel approach to confidential computing and
as such it opens many new research questions at the intersection of
cryptography and systems.
The most important theme among these questions is scalability and a problem we
refer to as the label management problem.
It is advantageous to refer to secret data in wire label form to avoid
evaluating a decryption algorithm under MPC for performance reasons, but the
wire labels cannot be reused in multiple circuits as that jeopardizes the
security of the garbled circuits.
When inputs to functions are sent as garbled circuit wire labels, how are the
labels generated using a secret shared across many remote clients and many
enclaves?
This not only extends to garbling generation but storage: how should labels be
stored across many untrusted hosts to keep the secret values consistent and
without reusing labels in multiple circuits?
Furthermore, how can functions be chained across machines when each are fed by
enclaves with different PRF seeds?
Can wire soldering protocols be used between circuits generated by different
enclaves~\cite{malishramprograms,lego,malish2pram,blackboxgram}?

\vspace{-2ex}
\section{Summary}
\label{sec:closing}
In this work we propose a method to evaluate short running, interactive
functions associated with FaaS platforms using confidential computing.
Our method moves function evaluation out of the enclave and onto the untrusted
host using our \shortsys{} protocol.
We motivated the need to pull the enclave out-of-band by showing that enclave
overhead for short running tasks is often greater than the task itself.
We then argued the security guarantees of doing so are the same as TEEs alone,
and lastly considered the performance implications.

\bibliographystyle{plain}
\bibliography{paper}

\end{document}